# Lifetime of a transient atmosphere produced by Lunar Volcanism

O. J. Tucker[1], R. M. Killen[1], R. E. Johnson[2], P. Saxena[1]

[1]NASA Goddard Space Flight Center, Greenbelt, Maryland 20771
[2]University of Virginia, Charlottesville, Virginia 48109

**Abstract.**

Early in the Moon's history volcanic outgassing may have produced a periodic millibar level atmosphere (Needham and Kring, 2017). We examined the relevant atmospheric escape processes and lifetime of such an atmosphere. Thermal escape rates were calculated as a function of atmospheric mass for a range of temperatures including the effect of the presence of a light constituent such as $H_2$. Photochemical escape and atmospheric sputtering were calculated using estimates of the higher EUV and plasma fluxes consistent with the early Sun. The often used surface Jeans calculation carried out in Vondrak (1974) is not applicable for the scale and composition of the atmosphere considered. We show that solar driven non-thermal escape can remove an early CO millibar level atmosphere on the order of ~1 Myr if the average exobase temperature is below ~ 350 – 400 K. However, if solar UV/EUV absorption heats the upper atmosphere to temperatures > ~ 400 K thermal escape increasingly dominates the loss rate, and we estimated a minimum lifetime of 100's of years considering energy limited escape.

1) **Introduction**

The possibility of harvesting water in support of manned space missions has reinvigorated interest about the inventory of volatiles on our Moon. It has a very tenuous atmosphere primarily composed of the noble gases with sporadic populations of other atoms and molecules. This rarefied envelope of gas, commonly referred to as an exosphere, is derived from the Moon's surface and subsurface (Killen & Ip, 1999). Outgassing from the interior and various space weathering processes maintain its surface pressure which is $\sim 10^{-15}$ times that of the Earth. Of great interest is the abundance of water in the form of gas, ice and frosts. Currently, water is delivered or formed during comet and meteoroid impacts, solar wind implantation with subsequent chemical bonding of H atoms, and outgassing of spacecraft missions. However, in the past a significant amount of magmatic water vapor might have been outgassed during episodic lunar eruptions.

Needham and Kring (2017), hereafter NK17, postulated that periodic lunar volcanism would have produced a transient atmosphere 3.8 – 1 Gyr ago. For the peak surface pressure 3.5 Gyr ago they estimated a lifetime of 70 million years using a thermal escape rate of 10 kg/s. This rate was inferred from an earlier study in which the atmosphere was assumed to be composed of atomic oxygen (16 amu) and possessed a much lower surface pressure, $10^{-17} - 10^{-10}$ bar (Vondrak, 1974). However, the NK17 surface pressures are on the order of mbar, similar to that of Mars, and their primary constituent was CO, mass 28 amu. Despite these differences, several recent studies have subsequently adopted the NK17 mass loss rate (Aleinov et al., 2019; Needham & Kring, 2017; Schulze-Makuch & Crawford, 2018; Wilson et al., 2019).

Here we consider the loss of the NK17 early transient atmospheres driven by atmospheric escape in detail for the first time. First, we review the applicability of the Vondrak escape rate to a millibar level atmosphere. Secondly, we comprehensively examine both thermal and non-thermal escape processes relevant for such atmospheres. Thermal escape is evaluated both using the theoretical Jeans escape equation with the exobase approximation for a single species atmosphere, and using a multispecies rarefied gas dynamic (RGD) model that accounts for outflow and gas rarefaction effects. The non-thermal processes considered are solar wind sputtering and photodissociative induced escape, which are currently important loss processes for heavy molecules in Mars atmosphere. We provide a framework for estimating escape from an early Moon atmosphere which can be further constrained with future exploration and sample collection of the lunar surface to be carried out by the Artemis program.

## 2) Creation of a Lunar atmosphere

In this section, we review the hypothetical atmospheres discussed in Needham and Kring (2017). NK17 estimated the total volume of mare basalts from 9 basins using measurements of the topography and surface gravity by Clementine and the Lunar Orbiter Laser Altimeter, respectively. Based on the frequency of crater sizes within the 3 largest basins, Imbrium, Procellarum and Serenitatis, they constrained mare volcanism to a time period from 1-3.8 Gyr ago over 0.1 Gyr time increments (Figure 1a). They found that during peak activity (3.5 Gyr) volcanic effusion could account for $10^{16}$ kg of molecules released into the atmosphere at an eruption rate of 10 kg/s. This atmospheric mass is equivalent to a global atmosphere with a surface pressure of ~ 9 mbar, slightly higher than the ~6 mbar on Mars. Volcanic glasses collected during the Apollo 15 and 17 missions provide evidence that fractional amounts of CO, $H_2O$, $H_2$ and OH were released during the eruptions, e.g., NK17 Table 1 and references therein. Based on the Apollo measurements we estimated mixing fractions for CO, S, $H_2O$ and $H_2$ of 0.33 – 0.4, 0.29 – 0.65, 0.005 – 0.01 and 0.0004 – 0.02 for the minimum – maximum degassed content, respectively. The mixing ratios were obtained using the total mass reported in the supplementary material, e.g., table S3 of NK17.

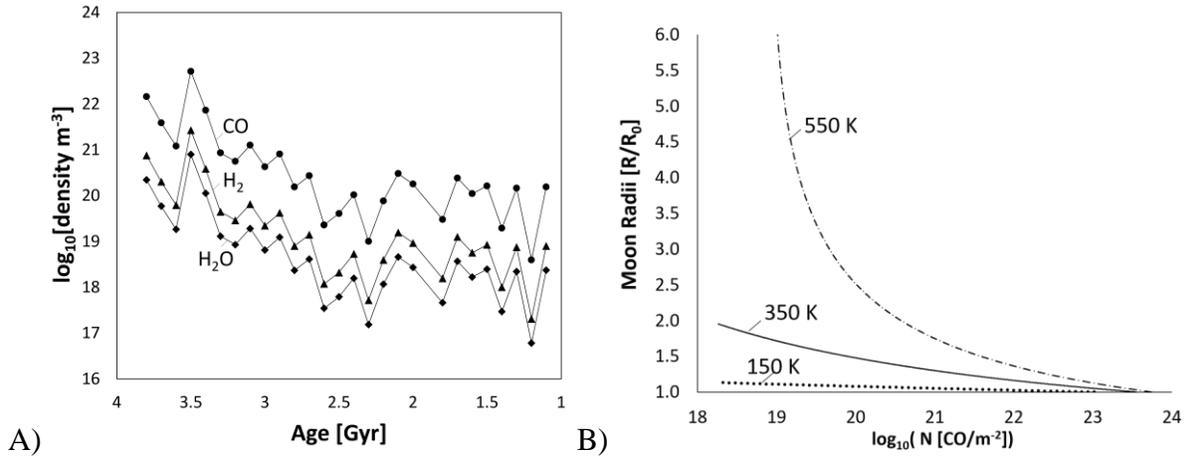

A)  B)

**Figure 1:** a) Near surface density as a function of lunar age calculated from the supplementary data of table 3 in NK17 under the assumption of an ideal gas. The circles, triangles and diamonds represent CO, $H_2$ and $H_2O$, respectively. b) Hydrostatic column density profiles for CO vs. altitude in lunar radii: $T = 100$ K (dotted curve), $T = 350$ K (solid curve) and $T = 550$ K (dash-dot curve)

The surface densities of outgassed molecules (CO, $H_2$ and $H_2O$) as a function of lunar age are shown in Figure 1a. The partial surface pressures of these species are given in the supplementary tables of NK17, and the surface density was calculated under the assumption of an ideal gas. We took an average of the maximum and minimum atmospheric masses for each of these species derived in NK17. The surface densities are dependent on an assumed depth of the outgassing source of lava, with other studies suggesting that the estimated surface densities are much lower if one does not use maximum impact basin depths as average depths (Wilson, Head, and Deutsch 2019). Additionally, the local magmatic lavas presumed to produce the global atmospheres may have possessed much higher temperatures ~1500 K (Stern, 1999), and that reservoirs of magma that produced lavas likely ranged in temperatures. Nevertheless, herein we adopted both the NK17 surface pressures and their assumed atmospheric scale height range, determined from the surface temperature, in order to consider the lifetime of the hypothetical atmosphere. As a baseline approximation, the atmospheric columns evaluated from the hydrostatic law indicate that these hypothetical transient atmospheres would be highly collisional. For example, the equation for hydrostatic balance applied to an isothermal spherically symmetric atmosphere is given below, where $H_a = kT_0/(GMm/rr_0)$.

$$n = n_0 \exp\{-(r - r_0)/H_a\} \qquad \text{Equation 1}$$

The corresponding atmospheric columns $\sim n_0 H_0$ are of order $\sim 10^{23} - 10^{28}$ CO/m² for surface densities of $\sim 10^{18} - 10^{23}$ CO/m³. In the preceding expression, the scale height at the surface is defined as $H_0 = kT_0/(GMm/r_0^2)$ with $r_0 = 1737$ km, $M = 7.35 \times 10^{22}$ kg and $m = 28$ amu defined as the Moon's radius and mass and molecular mass of CO, respectively. The other terms in the expression are $k$ the Boltzmann constant and $G$ the gravitational constant. Throughout the paper we refer to values of variables defined at the surface with the subscript 0. All of the transient surface pressures inferred in NK17 are consistent with the transient atmospheres being collisional, not ballistic. Further, for the predicted column densities, $> 10^{21}$ m⁻², the atmospheres are optically thick to solar UV/EUV radiation which, in turn, would drive the thermal structure of the upper atmosphere.

The mass loss rate used in NK17 was based on a study that examined the lifetime of volatiles that would be produced during human exploration of the Moon for an Apollo like mission (Vondrak, 1974). In particular, atomic oxygen was assumed to be the primary component of the artificial atmospheres and not CO, and the column densities of atomic O were orders of magnitude lower $\sim 10^{14} - 10^{21}$ O/m². Further thermal escape was assumed to dominate any non-thermal losses for columns larger than $10^{18}$ O/m². The thermal escape rates reported in that work were evaluated using the surface Jeans escape expression, discussed below. To this end, Vondrak (1974) applied an arbitrary heating rate as a function of the atmospheric column that increased the surface temperature to 800 K for columns larger than $10^{19}$ O/m². They calculated a mass loss rate of 60 kg s⁻¹ that was regarded to be independent of atmospheric mass for columns larger than $\sim 10^{19}$ O/m². NK17 adopted a loss rate of 10 kg s⁻¹ (Stern, 1999) that was more consistent with their derived volcanic outgassing rates. Based on the Vondrak study NK17 also assumed this loss rate is independent of atmospheric mass. However, as discussed in more detail below thermal escape must be evaluated considering both on the mass of the escaping molecule and the total atmospheric mass.

3) **Atmospheric Loss Processes Relevant for a Lunar Atmosphere**

This section reviews relevant thermal and non-thermal escape processes for the NK17 hypothetical atmospheres. First, we discuss the Vondrak (1974) escape calculation that has been subsequently applied to recent studies to show that the mass of the atmosphere must be considered when estimating escape in a collisional atmosphere. Secondly, we review common theoretical methods that are used to broadly estimate thermal evaporation, solar wind sputtering and photochemical escape from such atmospheres.

**3.1) Thermal Escape: Jeans Approximation, DSMC Modeling, Energy Limited Escape**

*Surface Jeans Escape*

First, we revisit the surface Jeans calculation carried out in Vondrak (1974). However, we evaluate the escape rate using the mass of CO instead of O. The thermal escape rate from a ***collisionless*** atmosphere in thermal equilibrium with its surface can be estimated by integrating the Maxwell Boltzmann speed distribution for speeds above the planetary escape speed $v_{esc} = (2GM/R_m)^{1/2}$. The expression given in Eq. 2 is often referred to as the surface Jeans escape flux, where $<v_0> = (8kT_0/\pi m)^{1/2}$ is the mean thermal speed and $\lambda_0 = GMm/r_0 kT_0$ is the Jeans parameter (J.H. Jeans, 1904; F. S. Johnson, 1971).

$$F_{J0} = \tfrac{1}{4} n_0 <v_0> \exp(-\lambda_0)(\lambda_0 + 1) \qquad \text{Equation 2}$$

For large Jeans parameters, $\lambda_0 > \sim 3$, escape occurs as collisions populate the fractional tail of the molecular speed distribution at speeds $v > v_{esc}$. However, as $\lambda_0 \to 1$ the mean thermal energy of particles approaches the planetary gravitational binding energy so that the bulk speed distribution leads to direct escape from the surface, $F_{J0} = \tfrac{1}{4} n<v_0>$ (Volkov, Tucker, et al., 2011). Since the escape flux ***exponentially depends on the molecular mass,*** for $T_0 = 300$ K, e.g., at the mean weighted lunar surface temperature, the O ($\lambda_0 = 18$) escape rate would be $\sim 10^6$ orders of magnitude larger than the CO ($\lambda_0 = 32$) escape rate. Even for the highest temperature of 800 K considered in Vondrak (1974), the surface Jeans rate for CO is 75 times smaller than that for O, as shown in Table 1. Therefore, although the Vondrak (1974) escape rate is reported as a mass loss rate it cannot be directly applied to the NK17 atmospheres which are composed of CO, S, $H_2O$ $H_2$ and OH.

**Table 1: Surface Jeans Escape Rate for O compared to CO**

| Mass[amu] | $\lambda_0(T_0 =800$ K$)$ | *$\varphi_{J0}$ [kg s$^{-1}$] |
|---|---|---|
| 28 | 12 | 0.8 |
| 16 | 7 | 60 |

Lower bound density taken from Vondrak (1974) atmospheric mass $10^7$ kg (surface density $n_0 = 10^{14}$ m$^{-3}$) or column density of $10^{19}$ O/m$^2$. $\varphi_{J0} = F_{J0}\, 4\pi\, r_0^2\, m$ is the surface Jeans escape rate and $\lambda_0$ is the Jeans parameter defined at the surface.

*Jeans Escape at Exobase*

As shown in Figure 1b the column densities of the hypothetical transient atmospheres indicate they are in the collisional regime. Unlike current lunar conditions, when considering thermal escape from a collisional atmosphere the Jeans equation (Eq. 3) must be evaluated at an altitude where collisions between molecules are infrequent (Chamberlain, 1963; Volkov, Tucker, et al., 2011). The transition of a gas from collisional to collisionless occurs over a range of altitudes where the thermosphere transitions into the exosphere, the so-called exobase at which the Jeans equation is often used to approximate the thermal escape rate. In the following expressions the subscript 'x' is used to refer to parameters defined at the exobase altitude. The nominal exobase $r_x$ is defined by the altitude where the mean free path between molecular collisions, $l_p \sim 1/n_x\sigma$, is equal to the atmospheric scale height, e.g., $l_p = H_x$ or $N_x = 1/\sigma$ where $\sigma$ is the collision cross-sectional area. For CO molecules we adopted the hard sphere collision cross section of $5.5 \times 10^{-19}$ m$^{-2}$ (Bird, 1994). In Eq. 3 the Jeans equation is written in terms of the exobase parameters.

$F_{JX} = \frac{1}{4}\, n_x \langle v_x \rangle \exp(-\lambda_x)(\lambda_x + 1).$            Equation 3

**Table 2: Exobase and Surface Jeans Escape Rates of CO**

| $T_x$ [K] | $r_x/r_M$ | $\varphi_{Jx}$ [kg/s] | $\varphi_{J0}$ [kg/s] | $\lambda_{Jx}$ |
|---|---|---|---|---|
| 200 | 1.8 | $9.6\times10^{-8}$ | $5.7\times10^{-8}$ | 26 |
| 250 | 2.1 | 0.025 | 0.0097 | 18 |
| 300 | 2.85 | 1.2 | 0.39 | 13 |
| 400 | n/a | n/a | 982 | n/a |

Subscript 'x' refers to exobase values of $r_x$ radial distance (altitude), $T_x$ temperature, $\lambda_x$ Jeans parameter and the Jeans escape rate $\varphi_{Jx} = F_{Jx}\, 4\pi\, r_x^2\, m$, $\varphi_{J0}$ is the surface Jeans rates of CO for comparison. The exobase values were derived using Eq. 1 with a surface density of $\sim 3.9 \times 10^{21}$ CO/m$^3$.

NK17 estimated the lifetime of the total mass of the hypothetical atmosphere without considering the dependence on molecular mass. However, the thermal escape of each atmospheric species will differ significantly depending on molecular mass, as discussed above. Therefore, we used surface densities derived from the CO atmospheric masses given in the NK17 supplementary tables, and not the total combined atmospheric masses. The Jeans escape rates given in Table 2 were evaluated using the spherical hydrostatic isothermal approximation (e.g., Eq.1). To compare the surface and exobase Jeans escape rates for the range of temperatures listed in Table 2 we used the surface density of $\sim 3.9 \times 10^{21}$ CO/m$^3$ at 3.7 Gyr. Consistent, with previous studies the Jeans rate evaluated at the exobase is roughly proportional to the surface Jeans rate scaled by the factor of $\sim r_x/r_0$ when the exobase us near the surface (R. E. Johnson et al., 2015). However, this scaling can significantly underestimate the escape rate for $r_x \gg r_0$. Small changes in the exobase location or in atmospheric temperature results in significant changes in the escape rate because of the exponential dependence. Therefore, it is important to highlight *that the escape rate is not independent of surface density or temperature when the exobase rises above the surface*. To this end, the Vondrak (1974) loss rate is not applicable to the conditions of the NK17 hypothetical atmosphere.

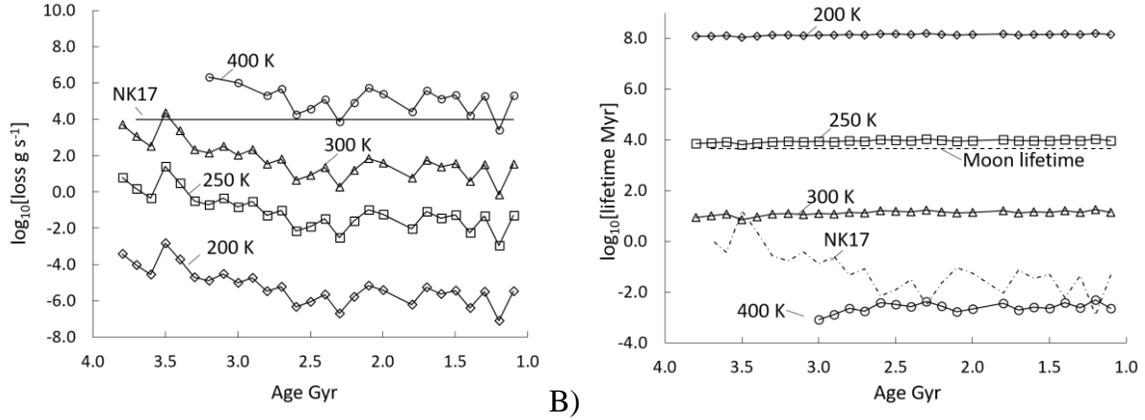

**Figure 2:** Left: Jeans escape rate for CO evaluated at the exobase using the surface density data from NK given in Fig. 1a assuming an isothermal atmosphere ($T_0 = T_x$ = 200 K, 250 K, 300 K, 400 K) versus lunar age (e.g. late lunar bombardment occurred 4.1 – 3.8 Gyr ago). The horizontal line labeled NK17 is the loss rate of 10 kg/s. Right: corresponding lifetimes for each transient atmosphere calculated using the loss rates in a). The dash-dot curve indicates the lifetimes obtained using the NK17 loss rate. The lifetime of the Moon is indicated by the dash curve.

In the following when referring to the 'Jeans rate' we refer to the exobase approximation and not the surface Jeans rate discussed earlier. In Figure 2a, the Jeans mass loss rates are plotted as a function of lunar age for each CO surface density given in NK17 where the isothermal hydrostatic equation was used to calculate the exobase altitude. The Jeans rates obtained for $T_0$ ~ 400 K roughly corresponding to the lunar subsolar temperature are only fortuitously consistent with the assumed NK17 loss rate of 10 kg/s, whereas at temperatures near 100 K roughly corresponding to the lunar night the loss rate is orders of magnitude smaller. Hence, again demonstrating the importance of accounting for atmospheric mass and temperature when estimating the thermal loss rates. In reality, the altitudes from which escape is probable must be considered carefully in the context of solar heating, radiative cooling and cooling via the expansion of the escaping gas. Global Circulation Models (GCM) have been used to investigate the atmospheric diurnal cycle driven both by the local surface temperature and atmospheric water content (Aleinov et al., 2019). Aleinov et al. (2019) obtained mean globally averaged temperatures of ~ 280 K for a dry CO atmosphere of ~10 mbar surface pressure. When including water with a surface mixing fraction of 0.005 they obtained much lower temperatures ~ 225 K due to radiative cooling. The globally averaged estimate of the escape rate shown in Figure 2a) is applicable to the Aleinov et al. model atmospheres which are firmly in the hydrostatic and nearly isothermal regime as discussed below.

We compare the atmospheric lifetimes obtained using the Jeans rate to that obtained using the NK17 loss rate of 10 kg/s for each inferred surface density in Figure 2b. The NK17 loss rate results in an atmospheric lifetime on the order of ~ 0.2 million years. This estimate assumes that the escape rate is independent of the surface density so the lifetime is primarily determined by the peak surface pressure at 3.5 Gyr. However, as stated above, the Jeans escape rate is a function of surface pressure, so it is lower for the time periods of lower surface densities because the exobase area is smaller and closer to the planet (lower $r_x$, larger $\lambda_{Jx}$) (Figure 2a). Therefore, unlike in NK17, for each 0.1 Gyr outgassing period when using the Jeans escape rate we find the lifetime of each transient atmosphere is relatively similar for a given temperature (Figure 2a & b). To this end, the volcanically derived CO atmospheres with an average exobase temperature < ~275 K are relatively stable against thermal escape over the Moon's lifetime. Note Aleinov et al. (2019) also suggested that with water vapor present in the atmosphere at temperatures warmer than ~175 K, CO would convert to $CO_2$. Hence, thermal escape is a much less viable loss process for an early $CO_2$ dominated atmosphere at current lunar temperatures.

*RGD modeling of multi-species atmospheres: adiabatic cooling and molecular diffusion*

In this section, we discuss gas kinetic simulations carried out to improve on the Jeans estimates of the thermal loss rate. We consider the escape of a multicomponent atmosphere to examine the effect of a small mixing fraction of a light escaping species, here $H_2$, on the heavy background species CO. Because escape causes adiabatic expansion and cooling of the upper atmosphere the inclusion of a light species can affect the heavier components and result in non-equilibrium molecule energy distributions near the exobase region due to infrequent collisions (Tucker et al., 2013). Models of the atmosphere using the hydrostatic approximation and diffusion equation do not account for non-equilibrium conditions, therefore solutions to the Boltzmann equation or gas-kinetic methods are needed. Here non-equilibrium is used to refer to molecular energy distributions that are not Maxwellian or when the gas temperatures of individual atmospheric species are different because of infrequent collisions.

Volkov et al., (2011) used a gas kinetic model to examine the breakdown of the thermal equilibrium in a single component atmosphere characterized by the source values of the Jeans parameter and the Knudsen number; $Kn_0 = (l_0/H_0)$ where $l_0$ is the mean free path for collisions. The Knudsen number characterizes the degree of rarefaction: i.e., small values $Kn \ll 1$ indicate the flow is in thermal equilibrium and large values $Kn > 1$ indicate rarefied flow. Typically the exobase is defined at the altitude where $Kn_x = 1$ (Chamberlain, 1963). However, thermal equilibrium models (hydrostatic equation, fluid equations, diffusion equation) can be very inaccurate at altitudes where the gas flow is increasingly rarefied, i.e. $Kn >\sim 0.1$. Despite this short coming, often thermal equilibrium models are used to approximate escape in this region of the atmosphere where collisions are infrequent. Therefore, using the results from gas-kinetic models we examine the conditions for which accounting for rarefaction effects in the NK17 atmospheres is important.

The range of surface pressures and atmospheric temperatures used in NK17 correspond to surface values of $Kn_0 \sim (10^{-9} - 10^{-3})$ and $\lambda_0 \sim (23 - 96)$, respectively. Applying the isothermal hydrostatic approximation with these surface pressures for temperatures higher than ~375 K an exobase is not obtained, e.g. Figure 3. This is the result of the isothermal hydrostatic approximation which asymptotically approaches a constant value of $n_0\exp[-\lambda_0]$ with altitude. If the Jeans parameter decreases to a value of $\lambda(r) = 2$ or $GMm/r = 2(kT)$ before an exobase is obtained, then the asymptotic density results in $Kn < 1$ at all altitudes. Therefore, the requirement for an exobase is not fulfilled at any altitude as indicated in Figure 3. In this scenario, the bulk thermal energy of the atmosphere is comparable to the planet gravity, and cooling due to the bulk outflow of the atmosphere must be considered. Gas kinetic simulations of the upper atmosphere can account for the effects outflow and rarefaction due to atmospheric escape.

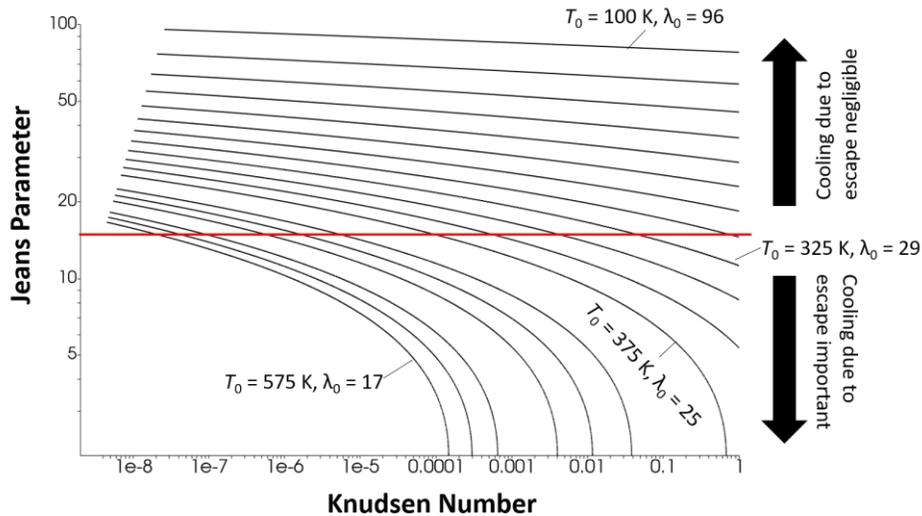

**Figure 3:** For a fixed atmospheric density at the surface, $n_0 \sim 3.9 \times 10^{21}$, the variation of the Jeans parameter as a function of altitude given by $Kn$ $(=l(r)/H_0)$ with $l(r)$ the mean free path for collisions vs altitude, $r$ with $Kn = 1$ the exobase altitude. The profiles are calculated for an isothermal, hydrostatic atmosphere with surface temperatures $T_0 = 100 - 575$ K. Profiles that do not intersect the left axis at $Kn = 1$ do not obtain an exobase. As indicated, cooling due to escape is important for density profiles below that intersect the red solid line.

Our previous gas-kinetic simulations of thermal escape of a single component atmosphere have shown that cooling driven by escape is important for atmospheres with Jeans parameters at the exobase of $\lambda(r_x) < \sim 15$ (Volkov, Johnson, et al., 2011; Volkov, Tucker, et al., 2011). Therefore, we used this criterion to identify the range temperatures for which the isothermal hydrostatic approximation should not be used. In Figure 3, the isothermal density profiles are plotted with black curves in terms of $Kn$ (density) and Jeans parameter (altitude) in the temperature range of $100 - 575$ K. Based on the Volkov et al. simulations adiabatic cooling due to outflow is important for all of the density profiles that intersect the red line. It is seen that the isothermal profiles with surface temperature $T_0 > \sim 325$ K ($\lambda_0 < \sim 29$) meet this condition. For density profiles that intersect the red line the escape rate would be significantly overestimated as cooling is ignored. We note Figure 3 implies isothermal profiles, but this condition is also applicable when the net effect of solar heating and radiative cooling are taken into account. That is, if radiative balance results in $\lambda(r) < \sim 15$ below the exobase the effect of adiabatic cooling on the thermal structure is important. Further, when escaping light gases are present, even as minor species (i.e., water products in the early lunar atmosphere), they can also contribute to cooling of a heavy background gas. Tucker et al., (2013) showed that $H_2$ escape from Titan reduces the scale height of the principal species, $N_2$ and $CH_4$, by removing thermal energy.

Outflow of a planetary atmosphere can be modeled by solving the fluid equations with appropriate boundary conditions determined from gas-kinetic models or solely using a gas-kinetic simulation (Erwin et al., 2013; Tucker et al., 2012). Here we focus on the gas-kinetic simulations that self consistently account for both outflow and rarefaction effects when tracking molecular escape. Here we focus on the gas-kinetic simulations that self consistently account for both outflow and rarefaction effects when tracking the escape atmospheric molecules. The simulations in this study were carried out using the Direct Simulation Monte Carlo (DSMC) method which is equivalent to solving the Boltzmann equation (Bird, 1994). In this method, a set of representative particles are tracked on a radial grid subject to mutual collisions and gravity. The source of molecules from the lower boundary is Maxwellian with zero bulk velocity and all species are presumed to have the same temperature. At the upper boundary molecules on escape trajectories were removed. All other molecules that traversed the upper boundary were specularly reflected back into the domain to reduce computational time. Using a free upper boundary and tracking molecules to the Moon's hill sphere would have very little influence on our results, slightly increasing the escape rate and decreasing densities at altitudes very far from the body less than 10 %. The computational domain is discretized in the radial direction using cell widths restricted to $1/3^{rd}$ of the local mean free path in the radial direction initially defined using the isothermal hydrostatic law. A weighted average among the species is used to estimate the mean free path to define the cell width. The trajectories of representative particles were integrated using the $2^{nd}$ order Runge-Kutta method. Within each cell the collision probabilities were evaluated using the variable hard sphere technique for elastic collisions (Bird, 1994). Each computational particle represents a statistical number of real molecules defined by a weight $W$. Here we used a variable particle weight for the multicomponent atmospheres as described in Bird (1994). We also simplify the content by including one representative heavy species, CO, and a small fraction of a light species, $H_2$. We defined the weight of the more abundant heavy species as $W_{CO}$ and the weight of the fractional species as $W_{H2}$. When a collision occurred between CO - $H_2$ the post collision velocity vector of $H_2$ was always updated and the post collision velocity vector of CO was updated with probability of $W_{CO}/W_{H2}$. This approach does not explicitly conserve energy before and after a collision, however on average energy is conserved (Bird, 1994). We verified this in our simulations as well.

**Table 3**: **Multicomponent Thermal Escape Rates**

| $T_0$ [K] | $\varphi(CO,H_2)$ [kg/s] | $\varphi_J(CO,H_2)$ [kg/s] | $\lambda_X$ | $r_x/r_M$ | $r_{LB}$ [km] | $Kn_{LB}$ | $\lambda_{LB}(CO,H_2)$ |
|---|---|---|---|---|---|---|---|
| 200 | n/a, 70 | $3.6\times10^{-8}$, 73 | 29.6, 2.1 | 1.7 | 2678 | 0.11 | 31, 2.2 |
| 300 | 1.6, 60 | 0.9, 60 | 11.5, 0.82 | 2.84 | 3670 | 0.038 | 15, 1.1 |
| 400 | 948, 57 | 876, 42 | 3.9, 0.28 | 12.8 | 5834 | 0.011 | 7.1, 0.51 |

$r_M$ is the lunar radius, $r_{LB}$ is the lower boundary altitude used in the DSMC simulations. The number densities used at the lower boundary are $n(r_{LB})_{CO} = 2\times10^{14}$ m$^{-3}$ and $n(r_{LB})_{H2} = 4\times10^{12}$ m$^{-3}$. $\lambda_x$ and $r_x$ are the exobase Jeans parameter and exobase radial distance, respectively obtained using the DSMC results of temperature and density. $\varphi$ is the escape rate obtained from the DSMC simulation, and $\varphi_J$ is the Jeans escape rate evaluated at the exobase using the DSMC results for density and temperature.

We carried out DSMC simulations for a spherically symmetric upper atmosphere with lower boundary $r_{LB}$, such that $n(r_{LB})_{CO} = 2 \times 10^{14}$ m$^{-3}$ and $n(r_{LB})_{H2} = 4 \times 10^{12}$ m$^{-3}$ representing a CO dominated atmosphere with a $H_2$ mixing fraction of 2 %. Three simulations were performed each with a lower boundary temperature $T_{LB}$ of 200 K, 300 K, and 400 K, respectively. The lower boundary, $r_{LB}$, was defined by the altitude where $n_{CO} = 2 \times 10^{14}$ m$^{-3}$ taken from the isothermal approximation given in Eq. 1 for each corresponding $T_{LB}$. This approximation was required because the atmospheric density at the surface and the exobase differ by many orders of magnitude and simulating the entire atmosphere using DSMC is too computationally intensive. To this end, at altitudes below $r_{LB}$ the density profile is taken to be isothermal, and at altitudes above DSMC was used to directly simulate the thermal structure accurately accounting for rare collisions and escape. Therefore, the results from the DSMC simulations are a lower limit of the effect of cooling due to escape, which is indicated by the temperature gradient at the lower boundary seen in Figure 4. A combined fluid-DSMC approach is needed to model the entire atmosphere for surface densities corresponding to Knudsen numbers of $Kn_0$ ($l_{mfp}/H$) << ~ 0.01 (e.g., Erwin et al., 2013; Tucker et al., 2012). However, because the conditions of the early lunar atmosphere are not well constrained implementing a detailed fluid-DSMC model including radiative heating and cooling is beyond the scope of this initial study. Here the goal is to use accurate simulations including CO and $H_2$ collisions to examine the effect of including a minor escaping species on the thermal structure of the upper atmosphere and escape of CO. As described below the simulated results can differ from the isothermal approximations discussed above.

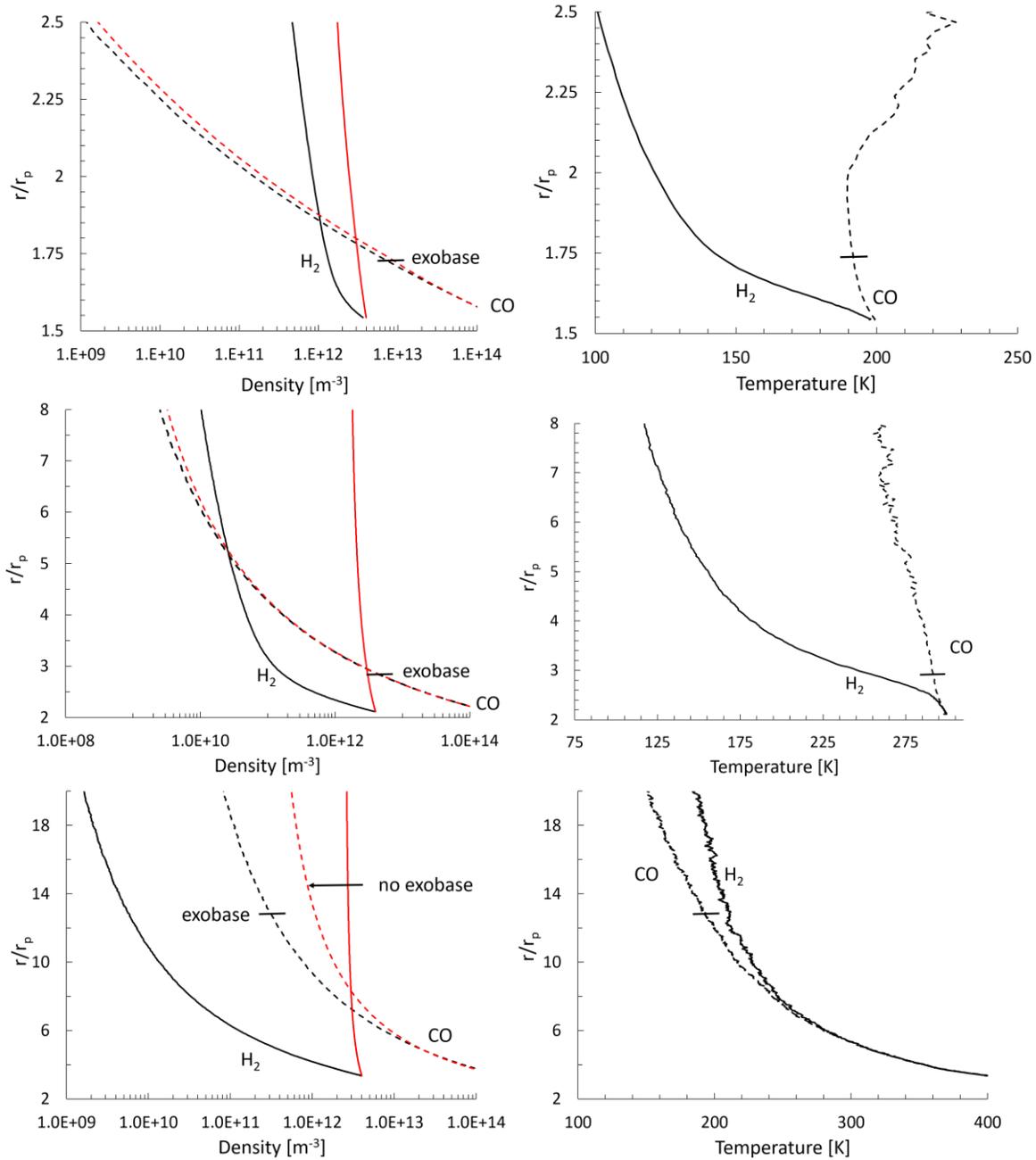

**Figure 4**: Comparison of DSMC results (black curves) and the isothermal hydrostatic approximation (red curves) for different lower boundary temperatures $T_{LB} = 200$ K (top panel), $T_{LB} = 300$K (middle panel) and $T_{LB} = 400$ K (bottom panel). The left panel shows the density profiles for CO (dashed curves) and $H_2$ (solid curves). The right panel only shows the DSMC temperature profiles for CO (dashed curves) and $H_2$ (solid curves) because the hydrostatic approximation is isothermal at all altitudes. The exobase for CO is indicated in each panel.

Hydrogen escape can have an important role in the early evolution of planetary atmospheres due to hydrodynamic escape (Lammer et al., 2018; Odert et al., 2018), and for current solar system bodies it can be an important cooling agent of the upper atmosphere (Mogan et al., 2020; Tucker et al., 2013). In the hypothetical early Moon atmosphere, $H_2$ escape is controlled by its diffusion through CO similar to that currently observed at Titan where the diffusion of $H_2$ is limited by $N_2$. Depending on the mixing fraction and atmospheric temperature, $H_2$ can dominate the gas distribution at high altitudes unless the temperature is high enough that CO escapes efficiently as well. For example, $H_2$ dominates at altitudes above 1.75 $R_m$ for $T_{LB}$ = 200, and 300 K, as shown in Figure 4. Whereas in the simulation with $T_{LB}$ = 400 K CO dominates the gas distribution at all altitudes. As shown by the DSMC results in Figure 4, adiabatic cooling results in both the density and temperature profiles differing from the isothermal hydrostatic approximation. In addition, it is seen that at altitudes below the exobase CO and $H_2$ have different temperatures indicating the gas is not in thermal equilibrium due to infrequent collisions. DSMC simulations with $T_{LB}$ = 200 K and 300 K not including $H_2$ would result in isothermal temperature profiles because CO does not escape efficiently at those temperatures, e.g., Table 3. However, when including $H_2$ its escape removes thermal energy from CO resulting in CO temperatures that differ from the isothermal approximation as shown in the top and middle panels of Figure 4. In the DSMC simulation with $T_{LB}$ = 400 K CO escapes at high rates resulting in much lower temperatures compared to the isothermal hydrostatic law. In addition, the DSMC results for density indicate an exobase altitude of 12 $R_M$ unlike this case for the isothermal hydrostatic law for which an exobase is not obtained.

DSMC more accurately estimates the escape compared to the exobase approximation which is applied at a single altitude assuming a Maxwellian gas. The Jeans escape rate for CO evaluated at the exobase of the DSMC density profile is slightly lower than that estimated for the isothermal profile because of $H_2$ cooling as discussed above. However, the actual escape rate is enhanced compared to the Jeans rates evaluated using either the DSMC or isothermal profiles $n(r)$ and $T(r)$ to define the nominal exobase. This is the case because the DSMC method accounts for molecules escaping from a range of altitudes and not just at the nominal exobase as in the Jeans approximation. Further, the DSMC results also highlight the ambiguity of the exobase approximation in a multi-species atmosphere because the species can have very different temperatures at the exobase. The $H_2$ mass loss rate is more than an order of magnitude larger than CO in the simulation with $T_{LB}$ = 300 K and several orders of magnitude larger in the simulation with $T_{LB}$ = 200 K. However, in the simulation with $T_{LB}$ = 400 K, CO which has a much larger surface source rate, also has a mass loss rate more than an order of magnitude larger than $H_2$. Fortuitously, the DSMC $H_2$ escape rate is relatively comparable to the Jeans rate evaluated using the CO temperature as shown in Table 3. For example, the $H_2$ Jeans rate is 40 % lower in the simulation with $T_{LB}$ = 200 K when evaluated at the CO exobase but using the $H_2$ temperature.

For completeness we compare our results of $H_2$ escape to the frequently used diffusion limited estimate in Hunten (1973). They proposed that the upward diffusive flux of $H_2$ can be equated to its escape flux when escape is 'easy' ($\lambda_0 \to <\sim 2$). In this approximation, the escape flux of the minor species depends on the scale height of the background gas, the binary diffusion parameter and its mixing fraction, which is assumed to be constant. Then the diffusion limited escape rate obtained from the binary radial diffusion equation neglecting the temperature gradient is given in Eq. 4,

$$F_{DL} = 4\pi r_0^2 \cdot b \cdot H_0^{-1} \cdot [1 - m_i/m_a] \cdot (1/f + 1)^{-1} \qquad \text{Equation 4}$$

where $b = 3\pi^{1/2}/(16 \cdot \sigma) \cdot (2kT_0/m_i)^{1/2}$ is the binary diffusion parameter, $m_i$ is the molecular mass of the minor species and $m_a$ is the molecular mass of the major species. The binary radial diffusion parameter is defined for a hard sphere gas and $\sigma$ is the collisional cross-section. The CO-$H_2$ cross-section used in the simulations discussed above is $\sigma = 4.7 \times 10^{-19}$ m$^{-2}$. Evaluating the diffusion limited flux equation using $T_{LB}$ = 200 K, 300 K and 400 K results in the mass loss rates of 58 kg/s, 47 kg/s and 41 kg/s compared to the DSMC escape rates of 70 kg/s, 60 kg/s and 57 kg/s, respectively. It is useful that the DSMC loss rates are only moderately enhanced relative to the diffusion limited values for the purpose of estimating the loss of minor species in early atmospheres because DSMC simulations are computationally demanding. However, the range of applicability of the diffusion limited approximation must be investigated in more detail.

Our DSMC results indicate that when escape is driven only by the surface temperature, the Jeans escape rate evaluated at the exobase underestimates the CO lifetime by a factor of 2, at most, for temperatures < ~ 325 K. However, at higher temperatures cooling due to escape must be accounted for when evaluating the Jeans rate at the exobase. Further, the results demonstrate that a single mass loss rate is not applicable to a multispecies atmosphere because the lighter species can be lost on much shorter timescales. Since escape driven by the surface temperature is not efficient and it significantly depends on the thermal structure of the upper atmosphere, below we consider heating of the upper atmosphere by the short wavelength solar radiation and then by the ambient plasma.

*UV/EUV Energy Limited Escape*

Since the temperature of the upper atmosphere on planetary bodies drives thermal escape, the absorption of solar UV and EUV, leading to molecular excitations, dissociations and ionization, can dominate the upper atmospheric heating (DeMore & Yung, 1999). For example, the upper atmospheres of Titan (~bar), Triton (~microbar) and Pluto (~microbar) are heated to temperatures above their respective surface temperatures by $CH_4$ (UV) and $N_2$ (EUV) photolysis. The hypothetical transient atmospheres considered here are more complementary to Mars. Although the current Mars' (~millibar) upper atmosphere is predominately heated by solar UV/EUV absorption of $CO_2$, $O_2$, CO and O (Fox & Dalgarno, 1979; Gu et al., 2020); thermal escape of these constituent is negligible over solar system timescales due to the large Jeans parameters. However, this is not the case at the Moon with its much lower gravity.

The possibility of the Moon losing a Mars' like atmosphere depends not only on the composition of the upper atmosphere but also how efficiently photochemical products transfer heat to the atmosphere. Escape of early planetary atmospheres is often simplified using the so-called energy limited (ELM) escape approximation (e.g., R. E. Johnson et al., 2015b and references therein), because of the complexity of accounting for a myriad of chemical and physical processes and the lack of detailed constraints on composition. Because thermal conduction in the thin region of an atmosphere is inefficient, energy limited escape assumes that the solar heating rate predominately goes into lifting atmospheric molecules out of the gravitational well, therefore the loss rate is roughly proportional to the heating rate. In this limit cooling by molecules through radiative emissions is *assumed* negligible. Therefore, a rough upper limit on the escape rate, given in Eq. 5, is often used, upper limit on the loss rate, where $r_a$ is the location of the EUV/UV absorption peak below the exobase but above the surface;

$dM/dt \sim mQ(r_a)/U(r_a)$, where $Q(r_a) = \eta 4\pi r_a^2 F_{EUV}$ Equation 5

here $Q(r_a)$ is the heating rate, $U(r_a)$ is the gravitational potential and d$M$/d$t$ is the resulting, globally averaged, mass loss rate. If the radiative cooling rate is small, a simple form of the heating rate as a function of $r_a$, heating efficiency η and the solar flux $F_{UV}$ can be derived from the fluid dynamics energy equation (Johnson et al., 2013). The ELM approximation is widely applied to estimate the loss rates from exoplanets (Lammer et al., 2018), early solar system atmospheres (Lammer et al., 2008) and Kuiper belt objects (R. E. Johnson et al., 2015). However, the limits of its applicability are still under consideration (Robert E. Johnson et al., 2013).

We used the ELM approximation to consider an upper bound estimate of the escape driven by EUV heating. Mars atmosphere is roughly optically thick to EUV for which the $CO_2$ and CO photoabsorption cross sections strongly depend on wavelength (DeMore & Yung, 1999; Heays et al., 2017). To this end we adopted an effective cross-section of the order of ~$10^{-21}$ m$^2$ for the ELM estimate. For example, observations of the Martian atmosphere reveal CO photodissociation peak at a distance of $r_a$ ~ 3520 km, where the column density is $N_a$ ~ $1.1 \times 10^{21}$ m$^{-2}$ (Cui et al., 2019; Fox & Dalgarno, 1979). Discussed below in section 3.2 we obtained a photodissociation peak for an equivalent CO column density when integrating the optical depth using the Heays et al. line specific photodissociation cross sections. The column density at the absorption peak can be correlated to the volumetric heating $Q(r_a)$. A radial column of ~$10^{21}$ CO m$^{-2}$ is obtained for the hypothetical lunar atmospheres in the range of ~1.2 – 2.1 $R_m$ (~2100 - 3700 km) using the hydrostatic law (Eq. 1) with $T_0$ = 300 K for the surface pressure in NK17. The EUV (89 – 107 nm) flux of the early sun was approximated by using the TIMED (Thermosphere Ionosphere Mesosphere Energetics and Dynamics) SEE (Solar EUV Experiment) solar irradiance level 3 spectrum obtained on an arbitrarily selected date of 2/9/2002. This flux is multiplied by an EUV enhancement factor, ~1.5 – 30, estimated for a more active early sun, 1 – 3.8 Gyr, given in Tu et al., (2015). Assuming a heating efficiency similar to that often used for Mars of ~ 16 % (Bougher & Dickinson, 1988; Fox & Dalgarno, 1979) the ELM loss rate is ~ $10^5$ – $10^7$ kg/s. Note this is more than ~4 – 6 orders of magnitude larger than the eruption source of 10 kg/s given in NK17. Likely, the estimated eruption rates are not proportional to the surface pressure as assumed in NK17 because much of the outgassed volatiles would have directly escaped the Moon. Nevertheless, the NK17 derived atmospheric masses would be lost on the order of ~ < 1000 years using the ELM escape rate. It is important to note that the ELM approximation is a rough upper limit on loss that has not been vetted by observation.

### 3.2) Solar Wind Sputtering and Photochemical Escape

Solar wind and pickup ions penetrating the atmosphere also induce atmospheric loss through non-thermal processes such as charge exchange, atmospheric stripping, and (sputtering) the direct ejection of neutrals via momentum transfer collisions (Johnson, 1994; Lammer et al. 2008). Recent analyses of lunar rocks have been used to infer that the Moon might have possessed a global magnetic field between the time periods 4.5 and 1 Gyr ago (Garrick-Bethell et al., 2019). For the maximum field strength considered, 5 µT, they derived a magnetopause distance of 3.7 $R_M$ which is on the order of or higher than the exobase distances of the hypothetical atmospheres considered with $T_0$ < ~325 K. Even if the Moon did not possess a dynamo after ~3.5 Gyr the solar wind interaction with the atmosphere and ionosphere would have induced a global ionopause similar to that observed at Mars, Venus and Titan (Nagy et al., 2004). In this situation, similar to Mars, pick-up ion sputtering might also be an important loss process (Jakosky et al., 2018; Leblanc et al., 2015). Detailed numerical simulations of the solar wind interaction with the early Moon atmosphere are needed to estimate the production and subsequent flux of pick-up ions onto the lunar atmosphere. Due to the lack of constraints on the early lunar electromagnetic environment, here we only consider the SW sputtering.

**Table 4: Solar wind Sputtering Yields**

| SW Ion | $Y_T$ | $S_n$ [eV cm$^{-2}$] | $\sigma_a$ [cm$^{-2}$] | $\sigma_d$ [cm$^{-2}$] | $<P_{ES}>$ |
|---|---|---|---|---|---|
| H$^+$ | 1.0 | $1.5 \times 10^{-15}$ | $18.3 \times 10^{-16}$ | $22.5 \times 10^{-16}$ | 0.5 |
| He$^+$ | 3.6 | $12 \times 10^{-15}$ | $24.3 \times 10^{-16}$ | $22.5 \times 10^{-16}$ | 0.5 |

*Estimated using an exobase distance of $r_x$ = 2.84 $R_m$ and $T_{LB}$ = 300 K in Table 3.

Atmospheric sputtering occurs when incident solar wind or magnetosphere ions penetrate near the exobase region and initiate a cascade of collisions in which atmospheric molecules obtain escaping trajectories. Johnson (1994) derived analytical expressions to estimate the number of ejected molecules per incident ion, referred to as the sputtering yield, using the Boltzmann transport equations. There are two components of the total sputtering yield $Y_T$ as indicated in Eqs. 6, molecules directly sputtered from single collisions $Y_{SC}$ Eq. 6a and molecules sputtered by recoil molecules produced in a cascade of collisions $Y_{CC}$ Eq. 6b. The single collision yield depends on the ion angle of incidence $\theta_A$, the ratio of the collision cross-section of the incident ion with atmospheric neutrals $\sigma_a$, the momentum transfer cross-section of atmospheric neutrals $\sigma_d$ and the mean probability for an atmospheric neutral to escape the column of atmosphere above the exobase.

$Y_{SC} = <P_{es}> \cdot \sigma_a/(\sigma_d \cdot \cos \theta_A)$  Equation 6a)

$Y_{CC} = (\beta/2) \cdot \alpha S_n/(\sigma_d \cdot U_x) \cdot 1/(\cos \theta_A)^P$  Equation 6b)

$Y_T = Y_{SC} + Y_{CC}$  Equation 6c)

where $\beta = 6/\pi^2$, $S_n$ is the stopping cross-section (eV cm$^2$), $\sigma_d$ momentum-transfer cross-section (cm$^2$) and the $U_x$ gravitational binding energy at the exobase (eV). Similar to other analytical sputter estimates discussed in Johnson (1994) studies we adopted an average value of $\cos \theta_A = 0.577$, and the values of the remaining constant parameters are $P = 1.6$ and $\alpha = 0.2$. $S_n$ and $\sigma_a$ depend on the energy of the incident ion and $\sigma_d$ depends on the thermal energy of atmospheric molecules. We obtained these values from the Atmospheric Escape Chemistry webpage on the NASA Planetary Data System, e.g., Table 4.

Garrick-Bethell et al. (2019) investigated the effect of a possible ancient lunar dynamo on SW precipitation to the lunar surface using a hybrid model to track SW protons. Stellar fluxes have been observed to decrease with the age, so they assumed an early solar wind flux of $16.5 \times 10^8$ H$^+$ cm$^{-2}$ s$^{-1}$ enhanced by a factor of ~6 of the current value. The estimated fraction of the SW flux reaching the surface was 9% and 33 % for averaged magnetic field strengths of 5 and 0.5 µT, respectively compared to the non-magnetized case. As a lower limit using these fractions for the early solar flux with the total yields of H$^+$ and He$^+$ (0.04 H$^+$) given in Table 4 we obtained atmospheric sputtering rates of 6 and 23 kg/s, respectively. Other studies estimate the SW flux could have been ~100 times larger 3.5 Gyrs ago (e.g., Wood et al., 2005). When using this higher flux we obtained an upper limit of total sputtering rate of 802 kg/s. In the above estimates, we approximated the interaction area with the solar wind protons at the exobase as $\pi r_x^2$ and $r_x = 2.84$ $R_m$ ($U_x = 0.3$ eV).

Photodissociation of CO and the subsequent escape of energetic C is also an important loss processes at Mars (Fox and Bakalian 2001). For example, carbon products are released with mean excess kinetic energy of ~1.5 eV (Cui et al. 2019; Fox and Bakalian 2001; Lee et al. 2014) which is comparable to their gravitational binding energy at Mars' surface. At the Moon this process is more efficient because the escape energy for C near the surface is much lower, ~ 0.4 eV. Therefore, at the Moon photodissociation is likely much more efficient than the electron-ion recombination process is a dominant loss process in Mars upper atmosphere (Lee et al. 2014).

The photo-dissociation rate ($PD$) at a given $r$ is a function of the line of site (LOS) optical depth ($\tau$) (Eq. 7a), and given by the integral in Eq. 7b where $F_\lambda$ is the solar photon flux (photons m$^{-2}$ s$^{-1}$ nm$^{-1}$). The LOS optical depth is approximated from the vertical column density using the Chapman grazing incidence integral $Chp$ (Smith & Smith, 1972).

$$\tau(r, \lambda) \sim N(r) \cdot Chp \cdot \sigma_{PA}(\lambda)/\cos(Z_A) \quad \text{Equation 7a)}$$

$$PD(r) = \int_{88\,nm}^{108\,nm} n(r)\sigma_{PD}(\lambda)F_\lambda e^{-\tau(r,\lambda)}d\lambda \quad \text{Equation 7b)}$$

Photodissociation of CO occurs via UV absorption in the wavelength range of ~88 – 108 nm (Heays et al., 2017). Adopting the photoabsorption $\sigma_{PA}$ and photodissociation $\sigma_{PD}$ cross-sections from Heays et al. (2017) we estimated the volumetric photodissociation rate using the Beer Lamberts law (DeMore & Yung, 1999). The optical depth for the globally averaged atmosphere is defined using the hydrostatic approximation and $Z_A$ is the solar zenith angle, here taken to be 60°. To obtain a simple estimate, here we also used the TIMED SEE solar irradiance spectrum to approximate the EUV flux (Fig. 5B), which again is scaled by the EUV enhancement of the early sun given in Tu et al., (2015).

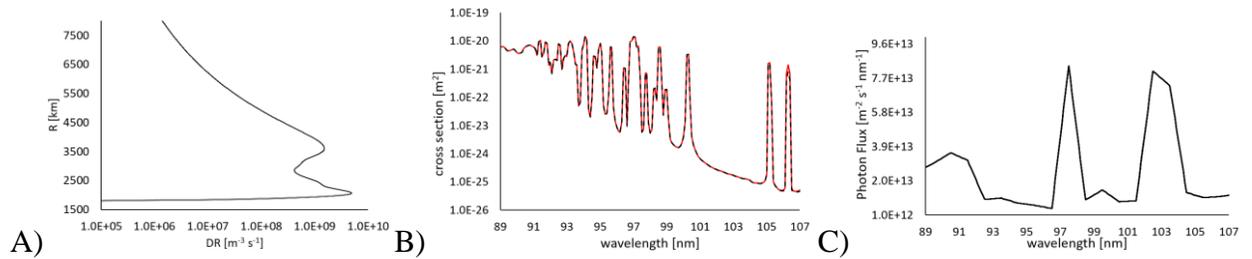

**Figure 5**: A) CO photodissociation rate vs. altitude for NK17 hypothetical lunar atmosphere at 3.7 Gyr using a hydrostatic column density, $N(r)$ with Eq. 7a, for $n_0 = 3.9 \times 10^{21}$ m$^{-3}$ and $T_0 = 300$ K. B)

Photoabsorption (dash black curve) and photodissociation (red solid curve) cross-sections (Heays et al., 2017) used to calculate the photodissociation rates (both curves overlap). C) Solar irradiance spectrum over the wavelength range of 89 – 107 nm.

An example altitude profile of the photodissociation rate of CO for the NK17 transient surface pressure derived for 3.7 Gyr is shown in Figure 5A. There are two peaks: one at $r \sim 3480$ km due to absorptions in the wavelength range of ~97 – 99 nm, and another at $r \sim 2030$ km due absorptions at longer wavelengths ~101 – 105 nm. The integrated destruction rate of CO down to the surface is $3.2 \times 10^{15}$ m$^{-2}$ s$^{-1}$ resulting in a significant production of carbon and oxygen radicals that can recombine to form other molecules. A detailed photo-chemical model is required to infer the steady state chemical equilibrium of atoms that do not readily escape, which is beyond the scope of this work. The mass loss rates due to atmospheric escape can be estimated using the column integrated photodissociation rates at the exobase. For a hydrostatic atmosphere with $T_0 = 300$ K and exobase distance of ~2.85 $R_M$ (e.g., Table 3) the mass loss rate is ~200 kg/s ($5.5 \times 10^{13}$ m$^{-2}$ s$^{-1}$) assuming 50% of dayside atoms escape using the hemispheric area ($1.5 \times 10^{14}$ m$^2$). Photochemical induced escape and solar wind sputtering are the dominant loss processes if upper the atmosphere temperatures are < ~400 K. In the absence of a dynamo, the photochemical loss rates are slightly lower than the sputtering rates, but within an order of magnitude. For magnetic field strengths ~5 µT (Garrick-Bethell et al., 2019), consistent with lunar samples, the estimated sputter losses are ~3 orders lower for peak surface pressures. The combined sputter and photochemical mass loss rates in the absence of a dynamo have a range of ~10 – 2000 kg/s corresponding to a collective atmospheric lifetime of ~1000's – 100000's years.

4) **Summary of Loss Processes and Implication for Lifetimes**

In this section, we discuss the implications of the thermal and non-thermal escape in the early lunar atmosphere. Figure 6, shown below, summarizes the mass loss rate and corresponding lifetime of each transient atmospheres proposed in NK17 for the relevant loss processes considered in this work.

*Thermal escape*: To estimate thermal escape we calculated the Jeans escape rate of CO at the exobase assuming an isothermal atmosphere (100 – 400 K) for each outgassing period.

Consistent with this assumption we showed that adiabatic cooling of due to CO escape is relatively negligible for temperatures below 375 K in section 3.1). Contrary to NK17 where the lifetime of the collective atmospheres is dominated by the outgassing period equated with the largest surface pressure (3.8 Gyr), we found the lifetime of each transient atmosphere to be comparable, within a factor of 2, when solely considering thermal escape.

When assuming an atmospheric temperature of $T_0 = 275$ K the lifetime of each transient is over ~200 Myr. Therefore, we can conclude that thermal escape is not a viable loss process for mean temperatures < 275K (e.g., Figure 6). Interestingly, Aleinov et al. (2019) examined the role of global heat transport for a lunar atmosphere with similar surface pressures, and they obtained average temperatures on the order of $T < {\sim}250$ K. They determined the early atmosphere would be dominantly heated by conduction from the surface, and that IR and UV absorption in the atmosphere is negligible. Further they suggested that CO could potentially convert to $CO_2$ and become the dominant volatile. Losing a millibar $CO_2$ atmosphere to thermal escape would require even higher temperatures > ~400 K. If CO or $CO_2$ were the dominant atmospheric constituents then condensation must have an important role as a loss process for the early atmospheres. Observing the composition of volatiles within the permanently shadowed regions would provide important constraints on the plausibility of a significant early lunar atmosphere.

Another important process is the drag-off of heavy minor constituents induced by the escape of a dominant lighter species, which has implications for the isotopic of lunar surface ices and regolith minerals. Such drag-off has been suggested to have fractionated heavy refractory elements like Cl, K, Na and Zn in lunar surface minerals early after the formation of the Moon (Paniello et al., 2012). Often the crossover mass, $m_{COM}$, is used to describe the minimum mass of a minor heavy species, $m_2$, which can be carried away by the upward force of the outflowing light minor species, $m_1$.

$$m_{COM} = m_1 + kTF_1/bgf_1 \qquad \text{Equation 8)}$$

If $m_2$ is less than the $m_{COM}$ it can be lifted of the gravitational well (Catling and Kasting 2017). Here $m_1$ is either CO or $H_2$, $F_1$ is the escape flux of the major species and $f_1$ is the mixing fraction $n_{CO, H2}/(n_{CO, H2} + n_2)$. The binary diffusion parameter $b$ was calculated using hard sphere cross sections of $5.5 \times 10^{-19}$ m$^2$ and $2.7 \times 10^{-19}$ m$^2$ ($b = 3.5 \times 10^{21}$ and $1.1 \times 10^{21}$ m$^{-1}$ s$^{-1}$), for CO and $H_2$ respectively. Table 5 gives estimates of $m_{COM}$ using Eq. 8 using the CO surface pressure of 0.16 mbar at 3.7 Gyr assuming a mixing fraction of $f_1 \sim 1$. It is seen that for $m_1$ as CO and $H_2$ drag-off is not a significant process when the upper atmosphere temperature is lower than 400 K. Of course, the analytical expression for $m_{COM}$ is a very rough approximation, so that rarified gas dynamic (RGD) should be carried out when possible as discussed below for a CO + $H_2$ atmosphere.

**Table 5: Estimates of crossover masses**

| *Escape Process | # Crossover mass (amu) | Mass Loss (kg/s) | $r_x$ [$R_M$] | T [K] |
|---|---|---|---|---|
| Jeans (CO) | 28 | $6.7 \times 10^{-4}$ – 1.6 | 1.6 – 2.84 | 300 |
| RGD (CO, $H_2$) | ~28.3, ~2.3 | 948, 57 | 12.8 | 400 |
| ELM (CO) | ~60.1 – 3306.1 | $10^5$ – $10^7$ | 1.2 – 2.1# | n/a |

#The crossover masses were evaluated using Eq. 8 with the exobase values from the RGD model and Jeans analytical results. For the energy limited escape (ELM) estimate the gravity at the heating peak $r_a$ is used.

The multispecies RGD can account for diffusion, escape and the thermal structure of the atmospheres self-consistently. We carried out such a simulation for a lunar CO atmosphere with a mixing fraction of 2% $H_2$ over a range of leading to escape rates of ~60 – 70 kg/s of $H_2$ which are roughly consistent with diffusion limited escape, Eq. 4. Below about 350K the $H_2$ escape rate dominates but only slightly cools CO. However, for temperatures > 400 K CO escape dominates the mass loss, and adiabatic cooling due to CO outflow is critical to take into account. When applying the RGD simulation with an upper atmosphere temperature of 400K we obtained an exobase distance of 12.8 lunar radii with a temperature < 200 K, whereas the isothermal approximation which does not obtain an exobase would grossly overestimate the thermal escape rate. For the range of surface pressures given in NK17 the hydrostatic isothermal Eq.1 is not applicable to estimate escape of CO for upper atmospheric temperatures > 350 K, e.g. Fig. 3.

As described above, the net UV/EUV heating of the upper atmosphere is the most important factor to estimate thermal escape and drag-off. To this end, the ELM escape approximation was used to estimate a rough upper bound to the thermal loss rate. When accounting for the higher EUV flux expected of the early Sun the atmosphere is short-lived, <

1000 years (Fig. 6). Estimates of the crossover masses for ELM escape indicate that such atmospheres could potentially drive severe mass fractionation (Table 5). However, the applicability of ELM is still in question (Robert E. Johnson et al., 2013; Salz et al., 2015). The ELM approximation can provide a reasonable estimate for atmospheres experiencing significant loss rates but with, counter intuitively, subsonic flow below the exobase (Johnson et al. 2013). For example, the ELM approximation is not be applicable to Mars present atmosphere in which EUV heating does not heat the upper atmosphere significantly enough to drive significant thermal escape of $CO_2$.

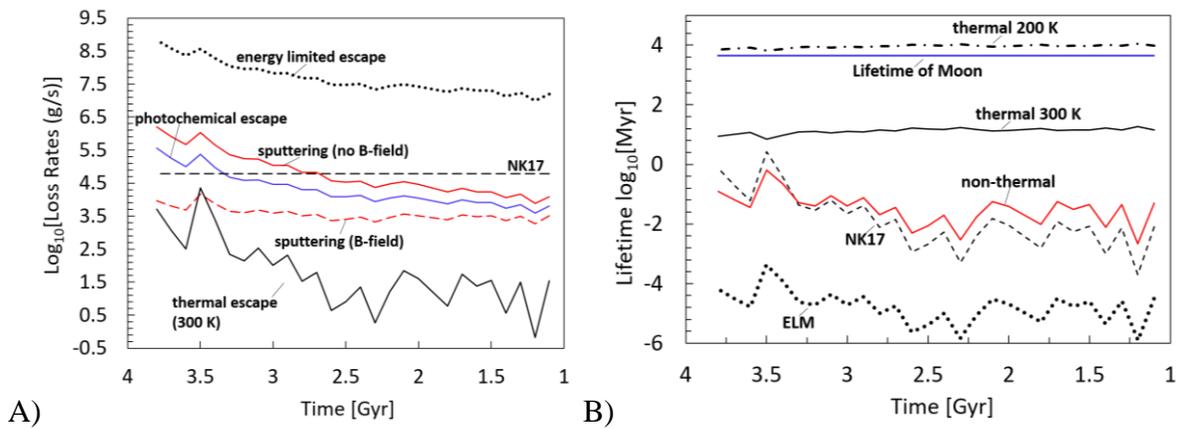

Figure 6: CO Mass loss rates and lifetime of transient lunar atmospheres (NK17) versus time for thermal and non-thermal (sputtering and photodissociation) escape processes.

*Non-thermal escape*: Both solar wind sputtering and photodissociation of CO are important loss processes. Without a magnetic field when accounting for losses due to sputtering and photochemical induced escape for the enhanced EUV/UV and plasma fluxes of the early sun we obtain similar CO mass loss rates as seen in Fig. 6a. In the presence of a an early magnetic field (Garrick-Bethell et al. 2019) the sputter loss rate is ~3 orders of magnitude lower. In addition, as the atmospheric column decreases over time solar UV/EUV heating is less efficient so that the photochemical loss rate decreases. The maximum combined non-thermal loss rates correspond to lifetimes of the hypothetical atmospheres ranging from 1000's of years to over 1 Myr and dominates the thermal escape for upper atmospheric temperature < 400 K. This finding is consistent with the current loss of $CO_2$ and CO occurring in Mars atmosphere. Lastly, sputter induced fractionation was not considered which could also contribute to mass fractionation.

The combination of future observations of volatiles in the PSRs and isotopic signatures of surface minerals combined with numerical experiments are important to decipher theories on the prevalent dynamics and evolution of volatiles in the lunar environment. For example, optical observations of the PSRs have been used to infer a water mass equivalent on the order of $10^{14}$ g (Eke et al., 2009; Hayne et al., 2015). Assuming the volcanic atmospheres were the source of hydrogen to the PSRs we can roughly estimate the accumulation rate required by using our estimated mass loss rates of $H_2$. The maximum $H_2$ mixing fraction given in NK17, $f = 0.02$ is consistent with a total mass ~ $10^{12}$ kg of $H_2$ in during the peak outgassing period. For an $H_2$ escape rate of 59 kg/s the lifetime of $H_2$ in the atmosphere against thermal escape is on the order of ~400 years. Based on this lifetime acquiring an inventory of $10^{11}$ kg within the PSRs requires an accumulation rate > ~ 8 kg/s. When considering the ELM estimates the corresponding $H_2$ loss rate is ~ 2000 – 2 ×$10^5$ kg/s, $f \cdot (10^5 – 10^7$ kg/s ), the atmospheric lifetime would be much less than 1 year requiring an accumulation rate more than 2 orders of magnitude larger.

5) **Conclusions**

NK17 estimated a total lifetime of 70 Myr for millibar level CO dominated atmospheres produce during periods of volcanic outgassing on the Moon. They adopted a mass loss rate of 10 kg/s presumed to be independent of atmospheric mass and attributed to thermal escape, consistent with Vondrak (1974). Unfortunately, this rate has been adopted in the other studies that consider the implications of the lifetime of volatiles in the early lunar atmospheres (Needham & Kring, 2017; Schulze-Makuch & Crawford, 2018; Wilson et al., 2019). However, as we have reviewed, the mass of the atmosphere must be considered in determining the mass loss rate.

The extent of thermal escape depends significantly on the assumed temperature of the upper atmosphere at altitudes where escape is probable e.g., $Kn > 0.1$. For temperatures < 400 K thermal escape of CO can be calculated using Jeans escape equation, Eq. 3, with the exobase approximation $Kn = 1$ (Chamberlain, 1963). The escape of minor species, however, will depend on their diffusion through CO. Here we considered molecular hydrogen, the dominant outgassed hydrogen bearing molecule, and showed its escape rate can be estimated using the diffusion limited approximation, e.g., Eq. 4. However, the actual thermal escape rates can differ from the Jeans rate adiabatic cooling, as shown using an RGD model, if the solar EUV/UV heating produces upper atmosphere temperatures > ~400 K and the isothermal law hydrostatic law in Eq.1 is not applicable. Lastly, an upper limit estimate of thermal escape obtained using the ELM approximation indicates that thermal escape could dominate the mass loss rate, although, this approximation is not well constrained.

Non-thermal escape driven by the higher UV/EUV and plasma fluxes of the early Sun, would dominate the mass loss rate for atmospheric temperatures < ~400 K. We calculated the photodissociation rates of CO using Eq. 7, and used the exobase approximation to estimate the amount of hot C and O lost to escape. Neglecting the past lunar magnetic field the sputter loss rate is comparable to the photochemical loss rates. However, when considering the Garrick-Bell paleo magnetic field the sputter loss rates can be more than 2 orders of magnitude lower.

Two important assumptions are used in this study are 1) the adopted surface pressures from NK17, and 2) temperature range considered for the hypothetical atmospheres. As more constraints become available on the surface pressure produced during the outgassing events the loss rates these estimates should be reconsidered with the appropriate atmospheric masses. In addition, detailed photochemical and solar heating models should be used to accurately estimate the thermal structure of the upper atmosphere. Subsequently, RGD models can use the resulting thermal conditions of the upper atmosphere to most accurately estimate escape.

**Acknowledgements**

O. Tucker and R. Killen acknowledge support of this work by SSERVI LEADER. The authors gratefully acknowledge Drs. Lynnae Quick (GSFC) and Parvathy Prem (APL) for useful discussions that contributed to this publication. O. Tucker dedicates this work to Richard A. Tucker Sr. (1933 - 2020) for his enduring encouragement.

https://doi.org/10.1029/1999RG900005

Tucker, O. J., Erwin, J. T., Deighan, J. I., Volkov, A. N., & Johnson, R. E. (2012). Thermally driven escape from Pluto's atmosphere: A combined fluid/kinetic model. *Icarus*, *217*(1), 408–415. https://doi.org/10.1016/j.icarus.2011.11.017

Tucker, O. J., Johnson, R. E., Deighan, J. I., & Volkov, A. N. (2013). Diffusion and thermal escape of H2 from Titan's atmosphere: Monte Carlo simulations. *Icarus*, *222*(1), 149–158. https://doi.org/10.1016/j.icarus.2012.10.016

Volkov, A. N., Johnson, R. E., Tucker, O. J., & Erwin, J. T. (2011). Thermally driven atmospheric escape: Transition from hydrodynamic to jeans escape. *Astrophysical Journal Letters*, *729*(2 PART II). https://doi.org/10.1088/2041-8205/729/2/L24

Volkov, A. N., Tucker, O. J., Erwin, J. T., & Johnson, R. E. (2011). Kinetic simulations of thermal escape from a single component atmosphere. *Physics of Fluids*, *23*(6). https://doi.org/10.1063/1.3592253

Vondrak, R. R. (1974). Creation of an artificial lunar atmosphere. *Nature*, *248*, 657–659.

Wilson, L., Head, J. W., & Deutsch, A. N. (2019). Volcanically-induced transient atmospheres on the Moon: assessment of duration and significance. In *Proceedings of the 50th Lunar Science Conference* (Vol. 50, Issue 1343).
34